\documentclass{nature_hack}

\usepackage[utf8]{inputenc}
\usepackage{float}
\usepackage{graphicx}
\usepackage[colorinlistoftodos]{todonotes}
\usepackage{amssymb}
\usepackage{longtable}
\usepackage{textcomp}

\usepackage{siunitx}	%comprehensive (SI) units package,eg:\SI{num(uncert)}{unit}
\DeclareSIUnit\torr{Torr}
\DeclareSIUnit\electron{e^{-}}
\usepackage{hyperref}

\title{Chemistry at graphene edges in the electron microscope} % Sure, why not

\author{Gregor T. Leuthner$^{1,}$\footnote{Email: gregor.leuthner@univie.ac.at}, Toma Susi$^{1}$, Clemens Mangler$^{1}$, Jannik C. Meyer$^{1,}$\footnote{Current address: Institute for Applied Physics, Eberhard Karls University of Tuebingen, Auf der Morgenstelle 10, D-72076 Tuebingen, Germany \& Natural and Medical Sciences Institute at the University of Tuebingen, Markwiesenstr.~55, D-72770, Reutlingen, Germany} \& Jani Kotakoski$^{1,}$\footnote{Email: jani.kotakoski@univie.ac.at}}

\begin{document}

\maketitle

\begin{affiliations}
 \item University of Vienna, Faculty of Physics, Boltzmanngasse 5, 1090 Vienna, Austria
\end{affiliations}

\begin{abstract}
    Transmission electron microscopy (TEM) and scanning TEM (STEM) are indispensable tools for materials characterization. However, during a typical (S)TEM experiment, the sample is subject to a number of effects that can change its atomic structure. Of these, perhaps the least discussed are chemical modifications due to the non-ideal vacuum around the sample. With single-layer graphene, we show that even at relatively low pressures typical for many instruments, these processes can have a significant impact on the sample structure. For example, pore growth becomes up to two orders of magnitude faster at a pressure of ca. \SI{e-6}{\milli\bar} as compared to ultra-high vacuum (UHV; \SI{e-10}{\milli\bar}). Even more remarkably, the presence of oxygen at the sample also changes the observed atomic structure: When imaged in UHV, nearly 90\% of the identifiable graphene edge configurations have the armchair structure, whereas armchair and zigzag structures are nearly equally likely to occur when the oxygen partial pressure in the column is higher. Our results both bring attention to the role of the often neglected vacuum composition of the microscope column, and show that control over it can allow atomic-scale tailoring of the specimen structure.
\end{abstract}

\newpage

% Introduction
Aberration-corrected~\cite{krivanek_electron_2008} scanning transmission electron microscopy (STEM) provides the ultimate spatial control for defect engineering~\cite{carr_defect_2010,sommer_electron-beam_2015} down to the level of individual atoms (see, for example Refs.~\citenum{susi_manipulating_2017,zhao_engineering_2017,mishra_single-atom_2017,jiang_atom-by-atom_2017} and references therein). However, the range of possible structural changes is normally limited by the elastic and inelastic interactions between the energetic imaging electrons and the sample to knock-on damage and radiolysis via electronic excitations~\cite{banhart_irradiation_1999,egerton_mechanisms_2012,susi_quantifying_2019}. While it is known that also chemical changes happen at the sample during microscopy experiments~\cite{leuthner_scanning_2019}, these are often neglected due to the uncontrolled composition of the residual vacuum. It thus remains unknown how large an impact residual vacuum gases can have on the sample during imaging. Additionally, if the atmosphere could be controlled, available parameters for defect engineering inside the microscope would expand from electron energy~\cite{meyer_accurate_2012}, dose rate~\cite{robertson_spatial_2012} and sample temperature~\cite{urita_situ_2005} to the vacuum level and composition~\cite{leuthner_scanning_2019}, ideally providing chemical control of materials down to the nanoscale.

Chemical effects have been reported for graphene nanopores~\cite{meyer_accurate_2012} via the comparison of pore growth rate at different electron acceleration voltages (20 and 80~kV), which ruled out the role of either knock-on damage or radiolysis or other processes arising from electronic excitations. Instead, structural changes at the pore edges were hypothesized to be caused by chemical processes through the interaction of the electron beam and molecules in the residual gases of the microscope column vacuum. Earlier transmission electron microscopy studies~\cite{girit_graphene_2009} have demonstrated a tendency for zigzag (ZZ) edges to form via sputtering under electron irradiation. This is in contrast to a recent environmental TEM study~\cite{thomsen_oxidation_2019}, where the formation of armchair (AC) edges was reported during high-temperature oxygen treatment at pressures in the \si{\milli\bar} regime, as well as the recent UHV STEM study where preference for AC edges was similarly reported~\cite{dyck_variable_2020}.

Graphene edges are also highly interesting in the context of graphene nanoribbons (GNR), for which their control remains a challenge in top-down production. GNRs have a width of just a few nanometers and electronic properties suitable for nanoelectronics applications~\cite{castro_neto_electronic_2009}. Although the competing bottom-up approach using organic precursor molecules has lead to impressive results~\cite{cai_atomically_2010}, it tends to be limited to metallic substrates. In the top-down alternative, graphene sheets are cut in the desired width and orientation. For applications, control over the exact edge structure of GNRs is critical since it directly affects their electronic properties~\cite{jaskolski_edge_2011,castro_neto_electronic_2009}: GNRs with edges in the AC crystallographic orientation are semiconducting, whereas those with ZZ edges are metallic. Previous etching experiments with GNRs have been carried out with scanning probe microscopy techniques on a substrate~\cite{wang_etching_2010,biro_nanopatterning_2010,oberhuber_anisotropic_2017}. For example, in Ref.~\citenum{wang_etching_2010} high temperature oxidation was used to produce sub-5-nm-wide GNRs from graphene on a Si/SiO$_2$ substrate that were imaged through non-atomic-resolution atomic force microscopy. However, no method has until now demonstrated control of different edge types at atomic resolution.

Here, we study the pore growth rate and atomic structure of edges in graphene as a function of the oxygen partial pressure in {\it in situ} STEM imaging experiments. Nanopores are initially created using a high accelerating voltage of 100~kV, whereas imaging is carried out at 60~kV to minimize direct knock-on damage. The pressure is controlled between near-UHV ($10^{-10}$~mbar) and $10^{-6}$~mbar using a leak valve at the microscope column, connected to a gas distribution line~\cite{leuthner_scanning_2019}. The residual gas composition is additionally measured with a mass spectrometer in the vacuum setup adjacent to the microscope column, connected through a flange facing the sample stage. We show that in near-UHV conditions, pore growth is practically nonexistent over typical experimental time scales, whereas it increases by orders of magnitude when oxygen is introduced into the column. Additionally, the atomic structure of nanopore edges---and more clearly of longer edges---depends on the oxygen atmosphere at the sample. At low pressures, AC edges are clearly preferred over ZZ (nearly 90\% of identifiable edges are AC), whereas the two edge types become equally likely at higher pressures.

\section*{Results and discussion}
\begin{figure}[b]
    \centering
        \includegraphics[width=\linewidth]{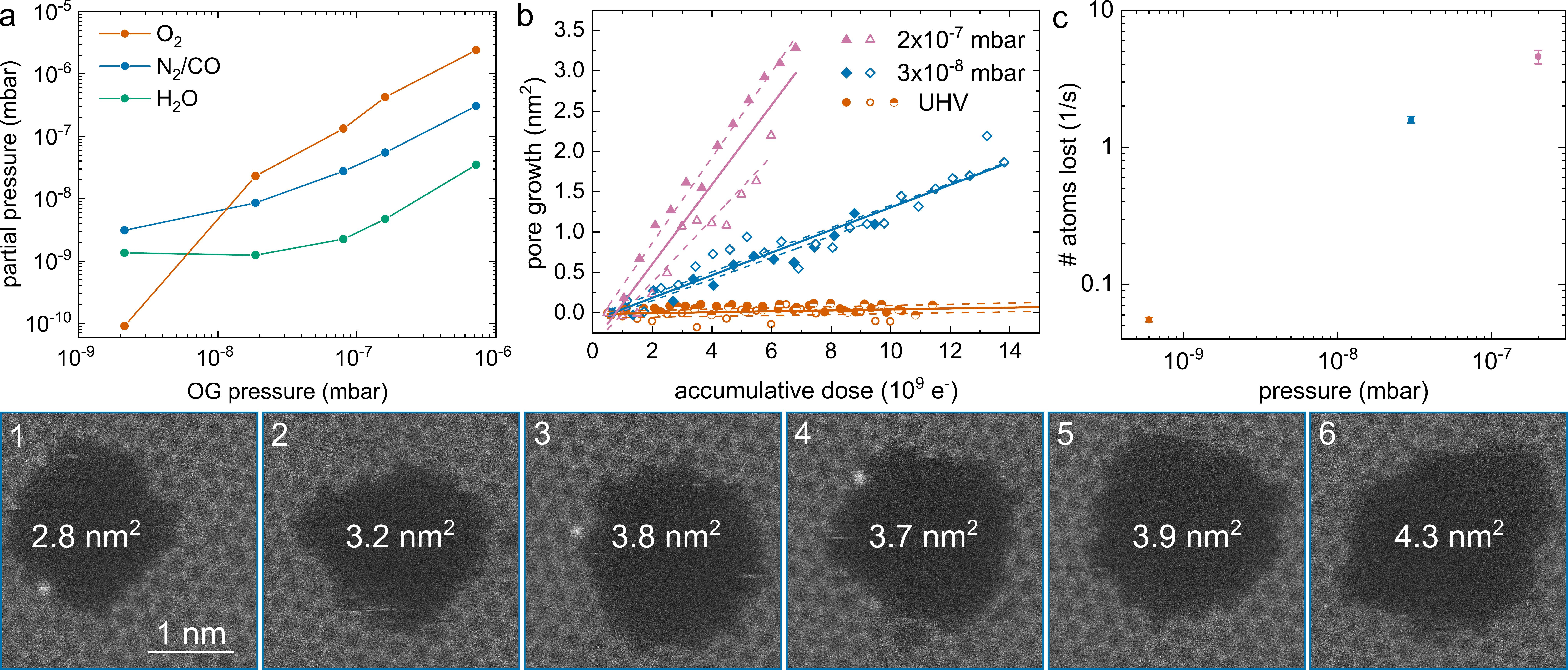}
    \caption{{\bf Pore growth rates at different pressures:} (a)~Partial pressures of three selected gases as a function of the objective gauge (OG) pressure. (b)~Pore growth for series at different pressures. For the UHV series, after 20 images only every 20th image (dose of $\sim$\num{1e10} electrons) was included in the analysis due to slow changes. (c)~Number of atoms lost per second as a function of pressure. (1)-(6)~Selected images of a \SI{3e-8}{\milli\bar} etching series.}\label{fig:poregrowth}
\end{figure}

We begin our experiments by creating nanopores into pristine graphene (commercial samples grown via chemical vapor deposition and transferred onto Quantifoil support grids by Graphenea Inc.). Since electron beam damage in pristine graphene is limited to knock-on damage~\cite{meyer_accurate_2012,susi_isotope_2016}, we use a relatively high acceleration voltage of 100~kV for this purpose. The electron beam is placed over a clean area (\SI{16}{\square\nano\meter}) of the sample, and irradiated for $\approx$~1~min with an approximate beam current of ca.~\SI{50}{\pico\ampere}. After this, the voltage is changed to 60~kV, the aberration corrector is retuned (the whole process takes about 30~min with the constant current mode of the Nion UltraSTEM 100 in Vienna; see Ref.~\citenum{dyck_variable_2020} for a description of a similar instrument). We first record image sequences of the created nanopores with a field of view of ca.~\SI{3}{\nano\meter} in near-UHV to establish a baseline for the pore growth rate. Next, we introduce oxygen into the column through a leak valve, and repeat the experiment for nanopore growth at higher pressures (up to $2\times 10^{-7}$~mbar). The measured gas composition for three gases with the highest partial pressures and the pore growth data are presented in Fig.~\ref{fig:poregrowth}.

As can be seen from Fig.~\ref{fig:poregrowth}a, the residual vacuum gas contains mainly N$_2$ (CO has the same mass, but there is little reason to assume it makes a significant contribution) and water with a trace amount of oxygen. Note that the data cannot reach down to the near-UHV pressure that is our normal condition, because the volume with the mass spectrometer has a base pressure of slightly above $10^{-9}$~mbar. When we start to introduce oxygen into the column, its partial pressure rapidly rises and it becomes the most prominent species of the residual vacuum. Overall, the partial pressure of O$_2$ is consistently an order of magnitude higher than that of chemically inert N$_2$ and two orders of magnitude higher than that of water. All reported pressures are readings from the objective area gauge of the microscope (OG), unless otherwise mentioned.

The etching rates corresponding to the growth of pores were calculated by measuring the pore area from image sequences at different pressures. This data, plotted against the cumulative electron dose, is shown in Fig.~\ref{fig:poregrowth}b with an example partial sequence of six images shown below. As can be immediately seen, pore growth in near-UHV is very slow (the complete image sequences were much longer than the data shown here, with no changes up to very high doses), while much higher rates are measured under oxygen atmosphere. Etching rates (calculated as the number of atoms lost per second) obtained by linear fits to the data are shown in Fig.~\ref{fig:poregrowth}c. This linearity indicates that the process is limited by the availability of oxygen atoms and not of etching sites. The difference in the etching rate between near-UHV (0.057~atoms/s) and $2\times 10^{-7}$~mbar (4.0~atoms/s) is nearly two orders of magnitude, highlighting the importance of the atmosphere on the structural changes in the sample during (S)TEM imaging. After the leak valve is closed, the vacuum recovers quickly (within half an hour) to the near-UHV values, bringing the etching rate back down to the values recorded before the leak valve was opened.

To put these numbers into context, let us consider an ideal gas at \SI{300}{\kelvin}. The impingement rate at a pressure of \SI{2e-7}{\milli\bar} is $\sim$\SI{1}{\per\square\nano\meter\per\second}. Assuming a mean residence time of $\sim$\SI{5e-10}{\second} (using an adsorption energy of \SI{0.16}{\electronvolt} for an O$_2$ molecule on graphene~\cite{zheng_density_2013}), the surface concentration of oxygen molecules should be $\sim$\SI{1e-9}{\per\square\nano\meter}, which is obviously too low to explain the observed etching rates. Note that even at the limit of the ballistic regime ($\sim$\SI{1e-4}{\milli\bar}), surface concentration would remain below $\sim$\SI{5e-7}{\per\square\nano\meter}. Such estimates are typically used to argue that the vacuum level in microscopes (often ca. $10^{-7}$~mbar) is sufficient to prohibit significant structural changes caused by chemical processes arising from the residual vacuum composition. As our results show, this is clearly incorrect. Moreover, the values we report here are consistent with recent literature. For example, analysing the data provided in Refs.~\citenum{meyer_accurate_2012,girit_graphene_2009} leads to etching rates of 0.15~atoms/s and 0.4~atoms/s, respectively, in non-UHV instruments with assumed vacuum conditions similar to our high-pressure experiments (the reported slowing of etching over time is consistent with the depletion of oxygen from the adjacent hydrocarbon-based contamination~\cite{leuthner_scanning_2019}). However, our results are consistent with the simple kinetic model of ideal gases if the relevant activation energy is $\gtrsim$0.7~eV. This is well below the adsorption energy of an individual O atom on graphene (ca. 0.92~eV~\cite{zheng_density_2013}), suggesting that a significant fraction of O$_2$ molecules close to the imaged area have been dissociated through radiolysis before landing on the sample.

We next turn to the detailed atomic structure of the pore edges during the experiments. Taking only knock-on damage into account, atomistic simulations~\cite{kotakoski_stability_2012} showed in 2012 that the removal of atoms from a graphene edge is determined by the dynamics of the edge atoms after they are hit by an electron, and not by equilibrium thermodynamics. Armchair edges were found to be more stable against such damage than zigzag edges due to their higher displacement threshold energy (energy required to displace an atom from its site in the structure). This is in contrast to the experimental findings from 2009 showing mostly zigzag edges under irradiation with \SI{80}{\kilo\electronvolt} electrons~\cite{girit_graphene_2009}, carried out at a typical vacuum level of $\lesssim$~\SI{e-7}{\milli\bar}. 

\begin{figure}[b]
    \centering
        \includegraphics[width=\linewidth]{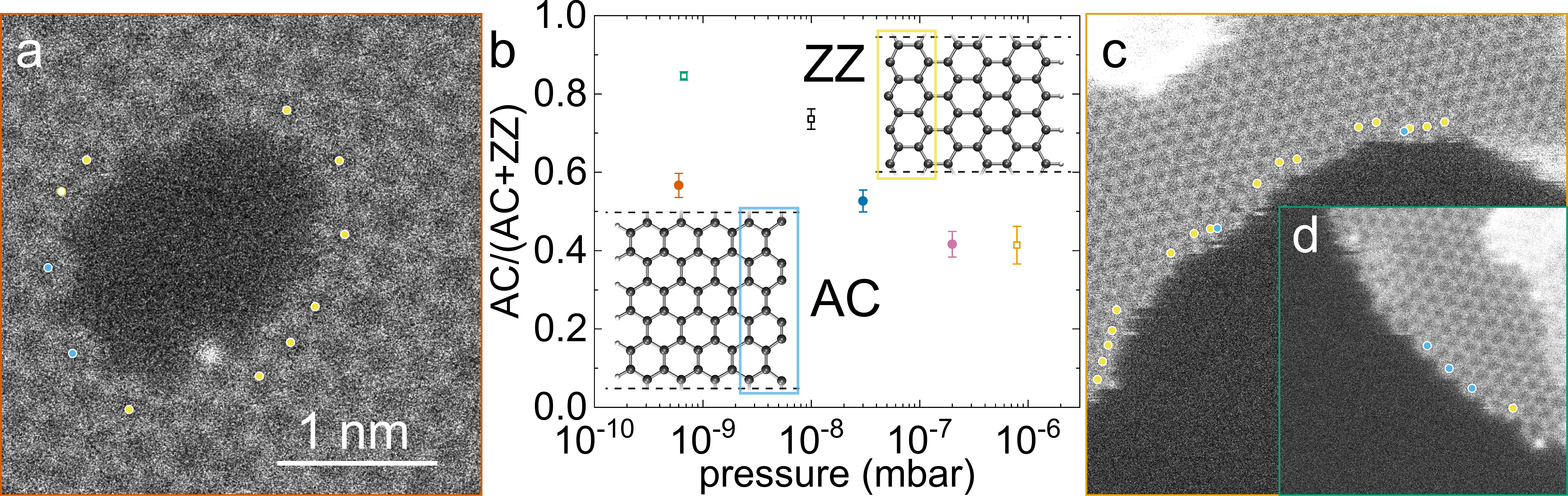}
    \caption{{\bf Armchair to zigzag graphene edge ratio:} STEM MAADF image (a)~of a pore and (c),~(d)~of graphene edges. Sky blue dots mark AC edge type cells, yellow ZZ ones. (b)~Ratio of AC/ZZ type edges normalized by unit cell length as a function of gauge pressure. Round (colored) data points correspond to the pore growth data and the square open symbols to data recorded at straight edges. For both cases data was collected at three different pressures, starting in near-UHV.}\label{fig:ACZZratio}
 \end{figure}

We classify the atomic structures at the pore edges in our images into five different categories: AC, most likely AC (may not be identified with total certainty due to local image contrast), ZZ, most likely ZZ, and ``other" (Fig.~\ref{fig:ACZZratio}a; see also Supplementary Information for examples of many of the ``other" configurations). We then compared the prevalence of AC and ZZ configurations by calculating the ratio of AC to all identified AC or ZZ configurations at each oxygen partial pressure. The results, normalized to the corresponding unit cell length, are shown in Fig.~\ref{fig:ACZZratio}b (full symbols). Although the results are quite similar at all pressures (ranging from ca. $0.56\pm 0.03$ in near-UHV to $0.42\pm 0.03$ at $2\times 10^{-7}$~mbar), there is a clear trend for the AC edges to become less prominent as the oxygen partial pressure increases.

As was also pointed out in Ref.~\citenum{kotakoski_stability_2012}, the circular shape of the pores may play a role in the observed difference between the AC and ZZ edges. It also complicates the analysis, leaving much of the edge structure into the category ``other". Therefore, we next turn to longer edges (for exemplary images, see Fig.~\ref{fig:ACZZratio}c,d) to avoid these complications. For these structures, the differences between the AC and ZZ edges become clearly more pronounced (see the open symbols in Fig.~\ref{fig:ACZZratio}b). Now, the near-UHV AC fraction is $0.85\pm 0.01$, whereas at $2\times 10^{-7}$~mbar it drops to $0.41\pm 0.05$. After the oxygen flow was stopped and the pressure was allowed to decrease down to $10^{-8}$~mbar, the fraction correspondingly increased to $0.74\pm 0.03$.

The prevalence of AC edges in near-UHV is easy to understand based on the simulation work of Ref.~\citenum{kotakoski_stability_2012}, as explained above. Nevertheless, to confirm this hypothesis, we also carried out the experiment for graphene grown with a 99\% $^{13}$C carbon source. As expected, all edges were now more stable in near-UHV with no marked difference between AC and ZZ edges, except for the transition from pure ZZ to the reconstructed ZZ-57 configuration~\cite{koskinen_self-passivating_2008} (see Supplementary Information). At higher pressures, the results were similar to those obtained with normal graphene, as expected for a chemical process with two isotopes of the same element.

In contrast to the AC edge stability, it is less clear why ZZ edges become more prevalent at higher pressures. A careful analysis of the images suggests that ZZ edges are particularly resistant against the chemical etching process. Indeed, we often notice that a ZZ edge remains stable over several images until one edge atom is removed. After this, the whole adjacent atomic row disappears. This is in contrast to the recent observation of higher stability of AC edges at higher oxygen pressures (up to \SI{6}{\milli\bar}) at elevated temperatures~\cite{thomsen_oxidation_2019}, pointing towards an explanation related to the absorption energetics of oxygen molecules or atoms on graphene or at its edges.

\begin{figure}[b]
    \centering
        \includegraphics[width=\linewidth]{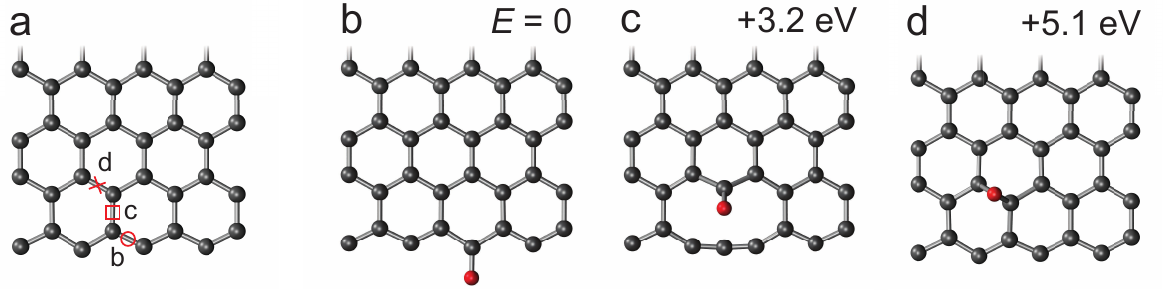}
    \caption{{\bf Density functional theory (DFT) modeling of zigzag edge oxidation.} (a)~Oxygen adatoms diffusing towards the edge may bond at C-C bridge sites at varying distances from it (circle, square, cross). (b)~Absorption of O at the bridge closest to the edge results in a barrierless reconstruction of the O to bind laterally to one edge C atom in the lowest energy state. (c)~Absorption on the next-nearest C-C bridge results in a barrierless reconstruction of a C chain, \SI{3.2}{\electronvolt} higher in energy $E$. (d)~Adsorption on the furthest C-C bridge results in a stable adatom configuration that is \SI{5.1}{\electronvolt} higher in energy than the lateral edge absorption site.}\label{fig:zzdft}
\end{figure}

To bring light to this question, we turn to density functional theory simulations (see Methods). Due to the large number of possible reactions, our aim here is only to provide plausible mechanisms in support of our experimental findings. First, to understand the diffusion of atomic oxygen towards the zigzag edge, we studied the stability of O adatoms at nearby bridge (epoxide-like bonding) adsorption sites (Figure~\ref{fig:zzdft}). Carbonyl-like bonding (oxygen bound to one carbon atom) at the edge is by far the energetically most favored bonding configuration, one that is also reached without a barrier if an O atom is placed on the bridge site next to the edge (red circle in Figure~\ref{fig:zzdft}a). The threshold energy for knocking out a C edge atom bound to the O is about \SI{10}{\electronvolt}, thus slightly weakening the zigzag edge against electron irradiation. Placing an O on the next-nearest bridge site (red square in Figure~\ref{fig:zzdft}a) results in another interesting barrierless reconstruction: a C chain partly detaches from the edge to allow out-of-plane carbonylic bonding for the O atom. Only the third-nearest bridge site (red cross in Figure~\ref{fig:zzdft}a) results in a stable adatom configuration. However, the relative energetics of the stable absorption sites clearly indicate the preference for O to bind directly to the edge. 

The situation for the armchair edge is significantly more complicated (see Supplementary Information for an illustration). Again, O adatoms preferentially bind in carbonyl-like bonding to one of the edge C atoms. This configuration can with a negligible energy barrier turn into an intermediate state that can be further modified by the addition of O bridge adatoms. This leads to a step-by-step unraveling of the edge structure, leading to configurations that can be easily sputtered by the electron beam. O atoms adsorbed on bridge sites near the edge result in no reconstructions such as that observed for the zigzag edge, and the energies of the different configurations are within a few hundred \si{\milli\electronvolt}. We further estimate that even with absorbed oxygen, the thermal dissociation of CO from the zigzag edge has a barrier of about \SI{4.0}{\electronvolt}, and about \SI{3.3}{\electronvolt} from the armchair edge. Thus both edges are expected to be stable under ambient conditions, and the electron beam must be involved in the etching process through a chemically assisted knock-on process.

\begin{figure}[b]
    \centering
        \includegraphics[width=\linewidth]{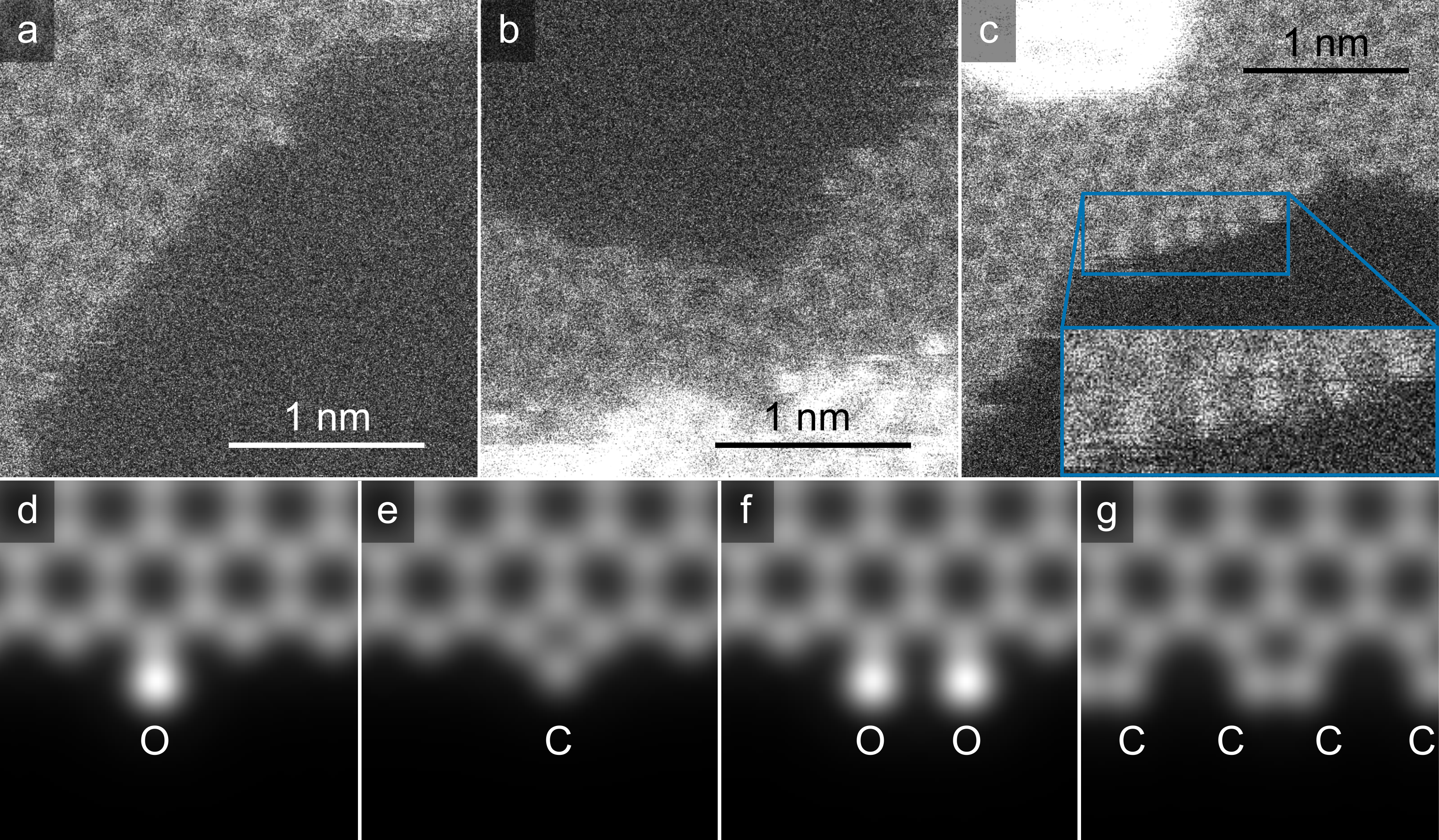}
    \caption{{\bf Atoms bound to graphene edges.} (a)-(c)~STEM-MAADF images of atoms attached to graphene edges. (d)-(g)~Image simulations of the energetically most favourable configurations obtained through DFT simulations when attaching a single or multiple carbon/oxygen atom(s) to a graphene ZZ edge. Oxygen forms Klein-like edges whereas carbon tends to dimerize.}\label{fig:klein}
\end{figure}

Careful analysis of the experimental images at high pressures appear to confirm the simulation results, at least for the ZZ edges. Indeed, we can see both carbon chains as well as individual oxygen atoms bound at the edges (similar to Fig.~\ref{fig:zzdft}c,b, respectively). Unfortunately, due to the electron-beam sensitivity of the structures, spectroscopic identification of the oxygen atoms proved impossible. To identify the atomic species protruding from the ZZ edges, DFT simulations were performed to obtain the energetically most favourable bonding sites for a single or for multiple carbon or oxygen atom(s). Figure~\ref{fig:klein} contains representative experimental images of atoms at the edge as well as multislice image simulations~\cite{madsen_abtem_2020} for DFT-relaxed configurations. The DFT simulations show that for oxygen, it is more favourable to bond in a Klein-like configuration (lower in energy by 1.60~eV, see Figure~\ref{fig:klein}d), whereas for carbon, bivalent bonding is preferred (by 0.45~eV, see Figure~\ref{fig:klein}e). Additionally, Klein edges seem to be only stable for oxygen atoms (Figure~\ref{fig:klein}f), as carbon edge atoms gain a significant amount of energy by dimerization (Figure~\ref{fig:klein}g). Comparing the simulated images to the experimental ones thus indicates that most of the atoms observed protruding from the edge should be oxygen, since such geometries do not correspond to ground-state configurations for carbon.

This conclusion is supported by intensity analysis. Our MAADF-STEM image simulations show that the intensity for a singly-coordinated carbon atom at a ZZ edge decreases compared to its neighbor (down to 92\%), whereas for oxygen it would increase (up to 158\%). In the experimental images (Figure~\ref{fig:klein}a-c), the intensities of the attached atoms are always higher than their neighbors (for some up to 120\%), which can not be explained by carbon atoms. These values, although lower than the simulated ones, are consistent with oxygen especially considering that the atoms at the edge tend to move during image acquisition, as seen from the streaked scan lines. 

We note that these results are not in contradiction with Ref.~\citenum{he_extended_2014}, because the TEM data provided there does not allow for elemental identification and was not recorded in UHV. In that study, both Klein edges and dimerized edges were observed, consistent with configurations containing both carbon and oxygen atoms. In Ref.~\citenum{suenaga_atom-by-atom_2010}, electron energy-loss near-edge spectroscopy was used to identify a single carbon atom in a Klein configuration, with no trace of oxygen. It is possible that the metastable carbon configuration has appeared due to strain or other influence of the surrounding structure. However, as noted above, we were also not able to detect oxygen signal at the edges, which we attribute to the high mobility of the atoms at the edge (in contrast to stationary oxygen atoms embedded inside the graphene basal plane, for which detection is possible~\cite{hofer_direct_2019}).

\section*{Conclusions} 
Electron-beam induced chemical processes potentially affect all materials, given a suitable composition of residual gases in the vacuum. Although these effects are typically neglected due to the unknown residual vacuum composition, they can play a crucial role in determining the stability of the sample in the microscope, but also affecting its atomic structure. In this work, we demonstrated that at oxygen partial pressures typical for non-UHV microscopes, the etching rate of graphene edges is up to two orders of magnitude faster than in near-UHV, where the process is so slow to be practically non-existent over experimental time scales. Additionally, while graphene armchair edges are significantly more resistant against knock-on damage due to their higher displacement threshold energy than zigzag edges, they are effectively destabilized under oxygen atmospheres. Thus there exists an active competition between physical and chemical processes in the electron microscope that can be tuned via acceleration voltage (high voltage leads to more elastic scattering and thus physical changes) and the sample atmosphere (chemical processes depend on the availability of suitable molecules). Due to the importance of atomically defined edges in applications with graphene nanoribbons, our results may also provide new ways to create nanostructured samples with unprecedented spatial resolution. Naturally, the possible material modifications through this method are not limited to etching but can be extended to any chemical manipulation where the end structure is sufficiently tolerant of electron irradiation. The method presented here demonstrates the possibilities of spatially controlled chemistry of materials, opening the way for a number of exciting new experiments. 

{\bf Data availability.} All the data generated or analysed in this study are included in this article and its Supplementary Information.

\bibliography{GTL_graphene_edges,references-3}

\section*{Methods}
Experiments were conducted with the aberration-corrected scanning transmission electron microscope Nion UltraSTEM~100 (Ref.~\citenum{krivanek_electron_2008}) at acceleration voltages of \SI{60}{\kilo\volt} and \SI{100}{\kilo\volt}. The convergence semi-angle of the beam was \SI{30}{\milli\radian}. The probe size of the instrument at \SI{60}{\kilo\volt} is on the order of \SI{1}{\angstrom}, which together with the scan area are important for limiting the electron beam effects to only the desired sample position. For acquiring images, a medium angle annular dark field (MAADF) detector with semi-angular range of \SIrange{80}{200}{\milli\radian} was used. The vacuum level in the sample chamber that can be reached with the instrument in Vienna~\cite{hotz_ultra-high_2016} is \SI{\sim 2e-10}{\milli\bar} (i.e., ultra-high vacuum, UHV). We point out that all pressures mentioned here are readings from the gauge in the objective area. The actual pressure at the sample is expected to be 5--6 times higher due to the geometry of the objective area of the microscope~\cite{leuthner_scanning_2019}. Additionally, it is likely that the gauge is less sensitive to O$_2$ than water molecules, which introduces another source of uncertainty.

The instrument is equipped with a leak valve system that adds the possibility to introduce gases into the sample chamber in the range of \SIrange[range-phrase=~to~]{e-9}{e-6}{\milli\bar} without affecting atomic resolution imaging~\cite{leuthner_scanning_2019} or causing noticeable changes in pressures measured in the neighboring volumes (i.e., the actual pressure remains below ballistic flow). The partial pressures of various gases as a function of pressure in the objective area when leaking in oxygen was measured with a mass spectrometer (Pfeiffer Prisma QME200). In the pressure range of the conducted experiments (\SIrange[range-phrase=~to~]{e-8}{e-6}{\milli\bar}) the measured oxygen partial pressure (by the mass spectrometer) starts slightly higher than the pressure displayed by the objective gauge. This discrepancy increases linearly to roughly half an order of magnitude (at \SI{e-6}{\milli\bar}). The partial pressure of other prevalent gases is always at least one order of magnitude (N$_2$/CO, Ar) lower than oxygen. The partial pressures of water, hydrogen and carbon dioxide are two orders of magnitude lower. Detailed results are shown in the Supplementary Information.

Unless otherwise specified, the samples were commercial freestanding monolayer graphene (Graphenea Inc.) with a natural isotope concentration grown via chemical vapor deposition on Quantifoil TEM grids. The samples were baked at~\SI{150}{\celsius} in a separate vacuum system or a chamber connected to the column for at least~\SI{10}{\hour} before being introduced into the microscope.

Our density functional theory (DFT) simulations were carried out with the GPAW package~\cite{enkovaara_electronic_2010} using the Perdew-Burke-Ernzerhof exchange-correlation functional~\cite{perdew_generalized_1996}. The model systems were six rows wide zigzag (48 C atoms and four passivating H atoms on one side of the ribbon) and five rows wide armchair (60 C and 6 H) graphene nanoribbons with 9 \textbf{k}-points along the periodic direction of the ribbon and at least \SI{10}{\angstrom} of lateral and vertical vacuum between the periodic images. Spin polarization was included in all calculations with a fixed total magnetic moment. After relaxing the structures~\cite{larsen_atomic_2017} with a force convergence criterion of 0.01~\SI{}{\electronvolt}/\SI{}{\angstrom}, we placed one or more atoms of oxygen or carbon at or near the ribbon edges to find the most favorable bonding sites, as well as any direct changes in the edge structure upon oxidation.

To further study the effect of oxidation on the electron-beam stability of the edges, we conducted molecular dynamics (MD) simulations to determine the displacement threshold energies of selected edge atoms following our established methodology~\cite{susi_atomistic_2012,susi_isotope_2016}. We considered in the simulations AC, ZZ and ZZ-57 edge configurations. The simulated displacement threshold energies for the configurations are, in agreement with earlier computational results~\cite{kotakoski_stability_2012}, \SIrange{18.75}{19.00}{\electronvolt} (AC), \SIrange{12.00}{12.50}{\electronvolt} (ZZ) and \SIrange{20.00}{20.25}{\electronvolt} (ZZ-57), and the pristine graphene value is \SIrange{21.75}{22.00}{\electronvolt} as calculated with the same method.

STEM image simulations were carried out using the multislice method within the abTEM package~\cite{madsen_abtem_2020}, with a spherical aberration coefficient of 1.5~\textmu{m}, a focal spread of 5~nm, and MAADF detector angles and illumination semiangle set to the experimental values.

\section*{Acknowledgments}
G.T.L. acknowledges the Uni:docs program of the University of Vienna and the Vienna Doctoral School in Physics for financial support. T.S. acknowledges funding from the European Research Council (ERC) under the European Union’s Horizon 2020 research and innovation programme via Grant agreement No.~756277-ATMEN, and C.M. and J.C.M. via 336453-PICOMAT. J.K. acknowledges support from the Austrian Science Fund (FWF) through project I3181-N36. Generous grants of computational resources from the Vienna Scientific Cluster are gratefully acknowledged.

\section*{Author contributions}
G.T.L. carried out most microscopy experiments and grew and prepared the $^{13}$C sample. T.S. carried out the DFT simulations. J.C.M. participated in the initial experiments. C.M. participated in the experiments and established the gas line for the experiments. J.K. participated in microscopy experiments, measured the residual gas composition with C.M., carried out the image simulations, supervised the study and carried out the analysis together with G.T.L. G.T.L. and J.K. wrote the article with contributions from all authors.

\section*{Competing interests}
The authors declare no competing interests.

\section*{Additional information}
{\bf Supplementary Information} contains results obtained with heavy graphene, additional observed atomic configurations at graphene edges, DFT modeling of oxygen etching at AC edges, mass spectrometer measurements of the vacuum composition and details about the edge type analysis. References~\citenum{krivanek_atom-by-atom_2010,koskinen_self-passivating_2008} are cited in the supplementary information. Additionally, original microscopy images for all shown figures are provided as supplemental material.\\
{\bf Correspondence and requests} for materials should be addressed to G.T.L. or J.K.

\newpage
\setcounter{figure}{0}

\renewcommand{\figurename}{Supplementary Figure}

\appendix

\section*{Supplementary information}

\subsection{Heavy graphene.}
To disentangle chemical processes from knock-on damage, additional experiments were conducted with CVD-graphene produced from $^{13}\text{C}$-enriched (99\%) methane, but otherwise similar to the commercial graphene samples. In the presence of oxygen (\SI{2.7e-7}{\milli\bar}), zigzag edges again clearly dominate, confirming the instability of armchair edges against atomic oxygen. However, the situation at low pressures differs from normal graphene (see Supplementary Figure~\ref{fig:c13maadfimages}) because all edge atoms are now considerably more stable against knock-on damage due to their higher mass. Indeed, hardly any atoms were removed from the edge during a 45~min experiment. Therefore, the edge is now much more constrained by the initial lattice orientation, although there still is a slight preference for armchair edges (Supplementary Figure~\ref{fig:c13maadfimages}a). When the initial edge is zigzag, however, during our experimental times no armchair edges were created. Instead, the edge can reconstruct into the zigzag-57 configuration\cite{koskinen_self-passivating_2008} (Supplementary Figure~\ref{fig:c13maadfimages}b), which has both a higher displacement threshold energy and lower formation energy than the zigzag edge. We also observed an edge with simultaneous AC and ZZ-57 configuration (Supplementary Figure~\ref{fig:c13maadfimages}c).

\begin{figure}[h!]
    \centering
        \includegraphics[width=\linewidth]{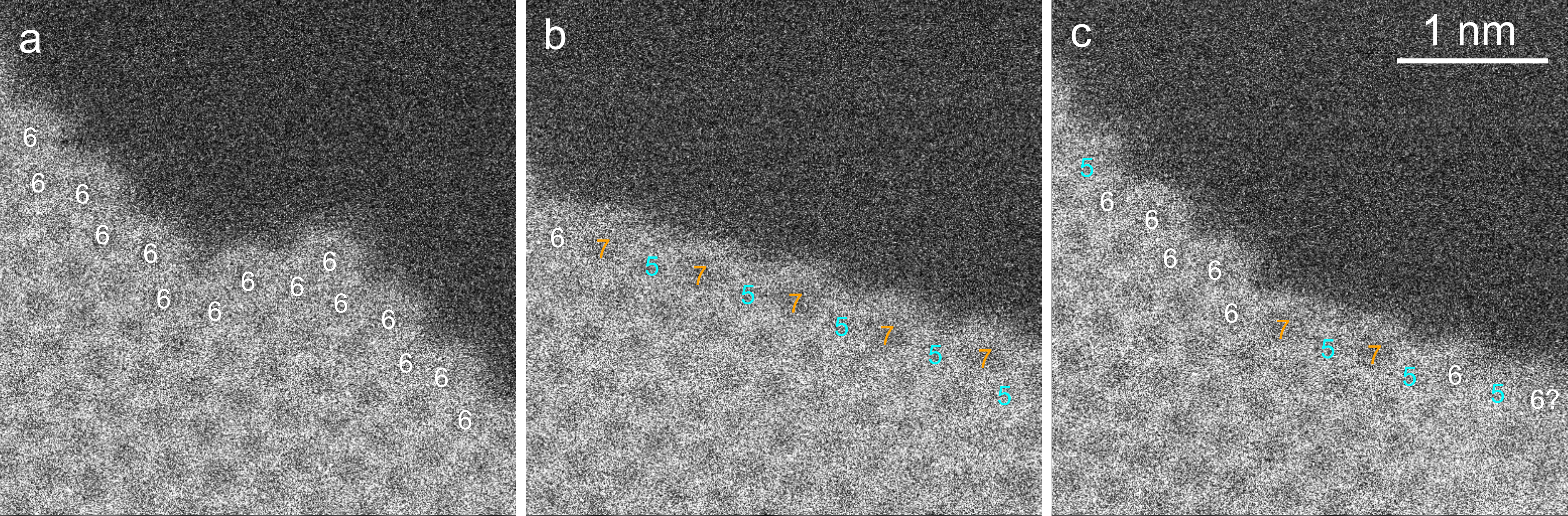}
    \caption{{\bf STEM MAADF images of $^{13}\text{C}$ graphene edges at a pressure of \SI{4e-10}{\milli\bar}.} (a)~Armchair configuration. (b)~ZZ-57 configuration. (c)~Transition of AC edge (left) to ZZ-57 (right). The overlaid labels show the number of atoms in each carbon ring.}\label{fig:c13maadfimages}
\end{figure}

 %\newpage
\subsection{Edge series in an oxygen atmosphere.}
In Supplementary Figure~\ref{fig:oxygenseries}, a series of consecutive STEM MAADF images of a graphene edge in an oxygen atmosphere of \SI{7.9e-7}{\milli\bar} is displayed. Various atomic arrangements have been marked to ease their identification and comparison to the atomic models obtained through simulations. These include plain zigzag (yellow) that is the most common case, followed by a carbonyl configuration (purple) where one oxygen atom (the brighter one) is bound to one carbon atom at the zigzag edge. This agrees with the DFT results that show carbonyl to be the lowest energy state for the oxygen atom at this edge. Also in agreement with the simulations, the formation of carbon chains (green) can be occasionally observed. The occurrence of plain armchair edges (blue box) is uncommon at higher oxygen pressures, but different states of armchair edge oxidation (blue circle) are sometimes captured. We also identified some furan-like (carbon pentagon with oxygen as heteroatom, orange circle) and ether-like (carbon hexagon with oxygen as heteroatom, brown box) configurations.

\begin{figure}
    \centering
        \includegraphics[width=1\linewidth]{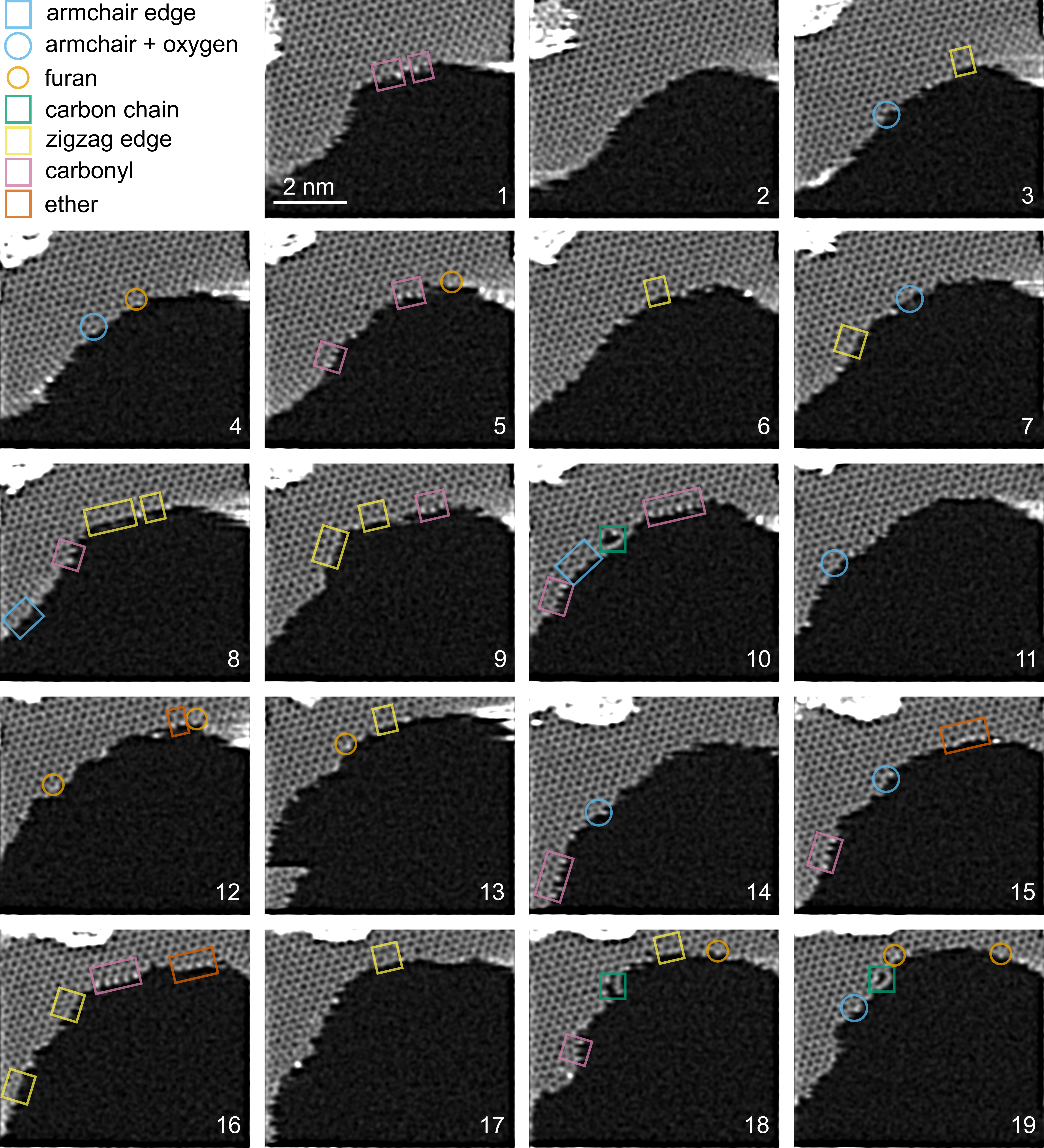}
    \caption{{\bf Series of STEM MAADF images of a graphene edge recorded at \SI{7.9e-7}{\milli\bar}.} The electron dose per frame is \SI[round-mode=figures,round-precision=2,scientific-notation=true]{2122113103.8599224}{\electron} and the total dose for this series is \SI[round-mode=figures,round-precision=2,scientific-notation=true]{40320148973.338524}{\electron}. All images have been treated with a double Gaussian filter\cite{krivanek_atom-by-atom_2010} to make the identification easier on a printout (raw data is provided as additional supplementary material).}\label{fig:oxygenseries}
\end{figure}

\subsection{Modeling of the AC edge.} 
\begin{figure}[hb]
    \centering
    \includegraphics[width=0.8\linewidth]{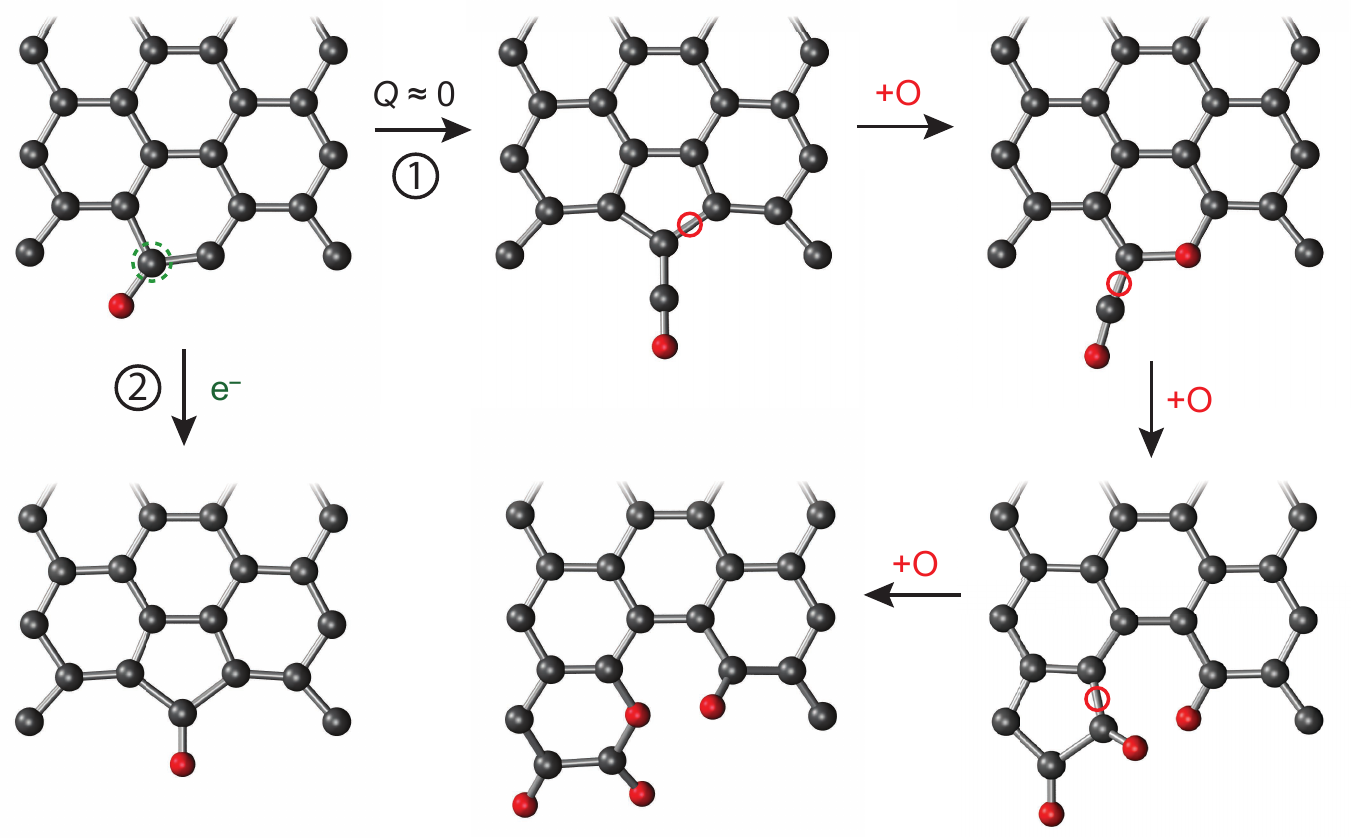}
    \caption{{\bf DFT modeling of armchair edge oxidation.} Due to the symmetry of the orbitals, the binding of a single O atom is not as favorable as in the ZZ case. Two distinct processes may follow: (1)~the edge atoms reconstruct with nearly no barrier $Q$ into a pentagon with a dangling C-O group. Further O adatoms may adsorb on the C-C bridges marked with red circles, spontaneously unraveling the edge; or (2)~if an electron transfers \SI{7}{\electronvolt} of kinetic energy to the C atom (green dashed circle) that the O is bonded to, it ejects the C from the structure, resulting in a pentagonal reconstruction that is relatively beam-stable. This configuration was rarely observed during the experiments, however, indicating that various processes similar to (1) are more likely.}\label{fig:acdft}
\end{figure}

DFT was used to study the unraveling of the AC edge (Supplementary Figure~\ref{fig:acdft}). Atomic O was placed on nearby bridge sites, leading to barrierless step-wise reconstruction into progressively more defective edge structures along chemical route 1. Alternatively, the C atom bonded to the initial O can be sputtered by the electron beam along physical route 2.

\subsection{Gas composition in the vacuum system}	
The complete gas composition as measured with a mass spectrometer in the chamber adjacent to the microscope column (connected through a flange facing the sample stage) is shown in Supplementary Figure~\ref{fig:massspec}.

\begin{figure}[hb]
	\centering
	\includegraphics[width=0.8\linewidth]{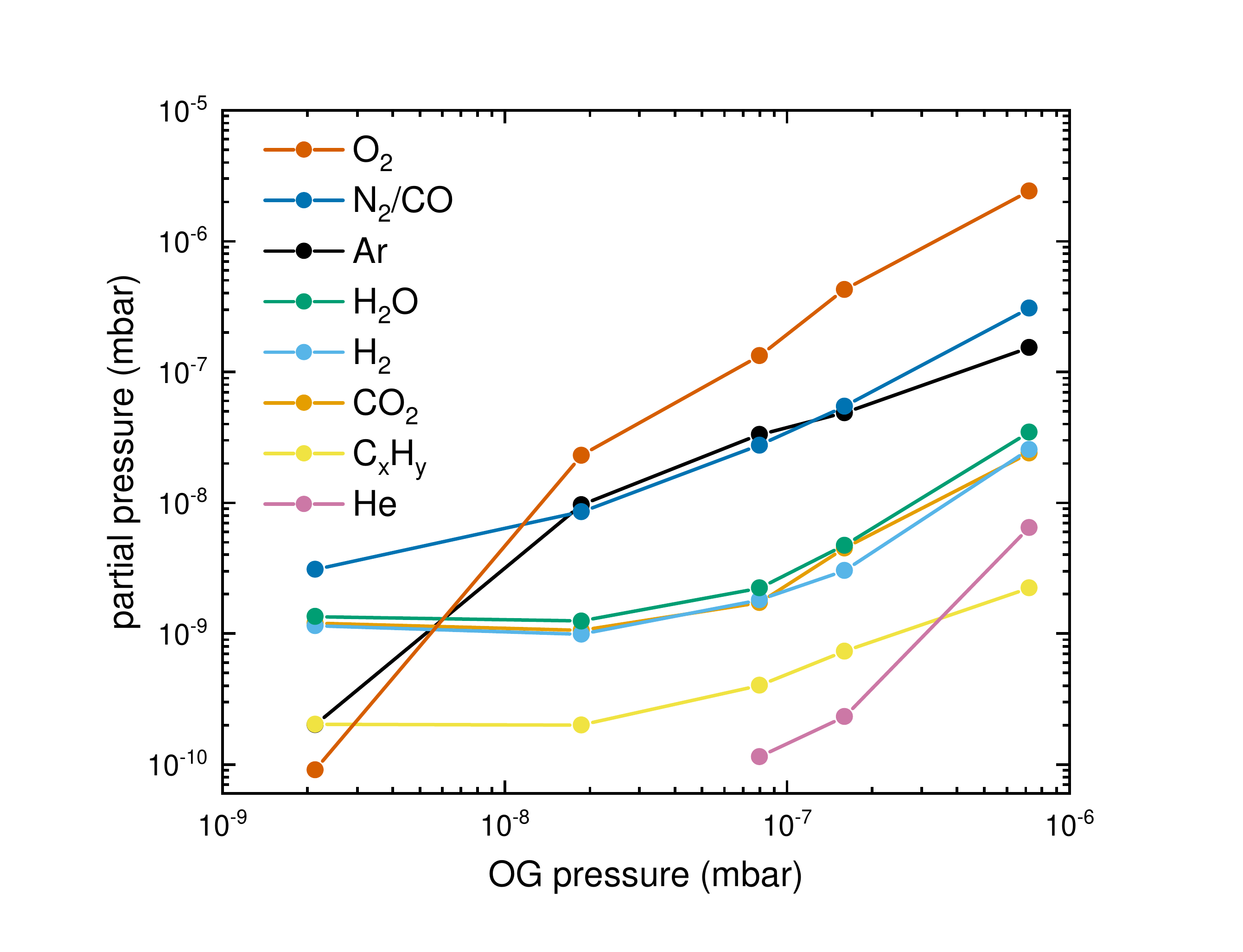}
	\caption{{\bf Gas composition.} Partial pressures of various gases when leaking oxygen into the objective area as a function of the objective gauge (OG) pressure.}\label{fig:massspec}
\end{figure}

\subsection{Edge type analysis}
\begin{figure}[t!]
	\centering
	\includegraphics[width=\linewidth]{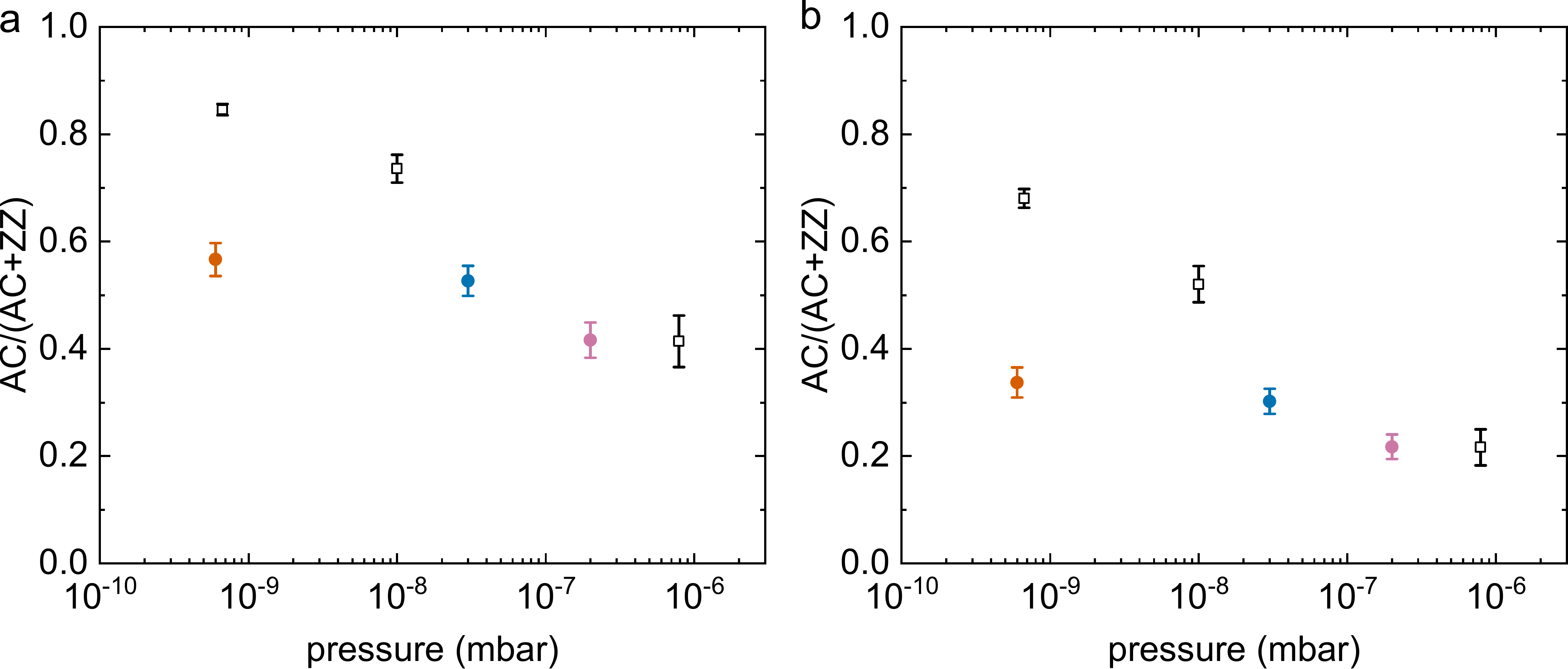}
	\caption{{\bf Edge armchair fraction.} (a)~Fraction with normalization to cell length. (b)~Fraction from number of cells without normalization.}\label{fig:ACZZnonnormalized}
\end{figure}

For the edge type analysis, AC and ZZ configurations along small pores and graphene edges were counted. To capture the etching dynamics, the scan speed was set to a relatively high value, especially for the pore growth data ($\sim$\SI{2}{\second} per image or $\sim$\SI{5e7}{\electron\per\square\nano\meter}). In series with high etching rates, the configuration change can happen during scanning a particular site. This can lead to stronger or weaker blurred spots, which can make it difficult to be certain about the exact atomic configuration. In these cases, AC or ZZ were marked differently and the number of these cases was included in the determination of the error bars. Volatile or heavier species at the edges (e.g.~moving Si atoms) were ignored in the analysis. The results were then normalized to the length of the respective edge type to ensure the comparability of different configurations at an edge with a given length. Supplementary Figure~\ref{fig:ACZZnonnormalized} depicts the comparison between normalized and non-normalized results, demonstrating that the conclusions do not depend on the normalization.

%In Tables~\ref{tbl:poreresults} and~\ref{tbl:edgeresults} are the end results by series. The in-detail results for every image analyzed are in Tables~\ref{tbl:porecells} and~\ref{tbl:edgecells} at the end of the supplement.

\end{document}